\newcommand{\be}{\begin{equation}}
\newcommand{\ee}{\end{equation}}
\def\Tr{{\rm Tr}}

\documentclass[
    ,final            
  ]
  {aipproc}
\setcitestyle{numbers}
\layoutstyle{6x9}

\begin{document}

\title{Signatures of phase transitions in nuclei at finite excitation energies}


\author{Y. Alhassid}{
  address={Center for Theoretical Physics, Sloane Physics Laboratory, Yale University, New Haven, CT 06520, USA}
}

\author{C. \"Ozen}{
  address={Faculty of Engineering and Natural Sciences, Kadir Has University, Istanbul 34083, Turkey}
}

\author{H. Nakada}{
 address={Department of Physics, Graduate School of Science,Chiba University, Inage, Chiba 263-8522, Japan}
}

\begin{abstract}
The mean-field approximation predicts pairing and shape phase transitions in nuclei as a function of temperature or excitation energy. However, in the finite nucleus the singularities of these phase transitions are smoothed out by quantal and thermal fluctuations. An interesting question is whether signatures of these transitions survive despite the large fluctuations.
The shell model Monte Carlo (SMMC) approach enables us to calculate  the statistical properties of nuclei beyond the mean-field approximation in model spaces that are many orders of magnitude larger than spaces that can be treated by conventional diagonalization methods. We have extended the SMMC method to heavy nuclei and used it to study the transition from vibrational (spherical) to rotational (deformed) nuclei in families of rare-earth isotopes. We have calculated collective enhancement factors of level densities as a function of excitation energy and found that the decay of the vibrational and rotational enhancements is well correlated with the pairing and shape phase transitions, respectively.
\end{abstract}

\maketitle


\section{Introduction}

A fundamental relation describing the thermodynamics of a system at temperature $T$ is the free energy
\be
F(T) = - T \ln Z(T) \;,
\ee
where $Z(T)$ is the partition function. For a system with a Hamiltonian $H$
\be
Z(T) = \Tr \,e^{- H/T} \;.
\ee
In the presence of correlations, $H$ is a many-body Hamiltonian, and it is difficult to calculate its partition function. A tractable approximation is the mean-field approximation, such as the Landau theory~\cite{Landau-book}. In the mean-field approximation, the free energy $F[T,\sigma]$ is evaluated as a function of certain order parameters $\sigma$ and then minimized with respect to $\sigma$ to find the equilibrium configuration. Examples of order parameters are the quadrupole deformation in the Landau theory of nuclear shape transitions~\cite{Alhassid1992}, and a complex pairing gap in the theory of superfluidity.

However, in a finite-size system such as the atomic nucleus it is often important to take into account fluctuations of the order parameters. In general there are two types of fluctuations: (i) thermal or static fluctuations, and (ii) quantal fluctuations. The probability for a static fluctuation $\sigma$ is  given by $P[\sigma] \propto e^{-F[T,\sigma]/T}$, and the partition function is approximated by
\be
Z(T) = \int D[\sigma] e^{-F[T,\sigma]/T} \;.
\ee
Quantal fluctuations become important at low temperatures and are included by considering time-dependent fluctuations $\sigma=\sigma(\tau)$ of the order parameters.

A systematic approach to include fluctuations beyond the mean-field approximation is based on the Hubbard-Stratonovich (HS) transformation~\cite{HS-trans}, in which the imaginary-time many-body propagator is described as a functional integral over one-body propagators of particles moving in external time-dependent auxiliary fields.  The saddle-point approximation leads to a mean-field theory. In the static path approximation (SPA)~\cite{SPA}, large-amplitude fluctuations of the static fields are integrated over, while in the SPA plus random phase approximation (RPA), small-amplitude time-dependent fluctuations are taken into account around each static fluctuation~\cite{Lauritzen1990,Attias1995}. In the auxiliary-field  Monte Carlo (AFMC) approach, also known in the context of the nuclear shell model as the shell model Monte Carlo (SMMC) method~\cite{Lang1993,Alhassid1994}, all fluctuation are included by a Metropolis sampling of the auxiliary-field configurations.

In a mean-field approximation the nucleus is predicted to undergo pairing phase transition and shape phase transition as a function of temperature or excitation energy.  Shape transitions are also predicted as a function of proton or neutron number. The fluctuations smooth the singularities of the phase transitions and an interesting question is whether signatures of these transitions survive despite the large fluctuations.

In the following we discuss the SMMC method and use it to study the emergence of collectivity in the configuration-interaction shell model approach~\cite{Alhassid2008,Ozen2012}. In particular, we focus on rotational collectivity associated with deformed nuclei and vibrational collectivity associate with spherical nuclei. We identify a thermal observable that can distinguish between the different types of collectivity and use it to study the transition from spherical to deformed nuclei in families of rare-earth isotopes. We define a collective enhancement factor of level densities and demonstrate that the decay of vibrational and rotational collectivity is correlated with the pairing and shape transitions, respectively.

\section{The Shell model Monte Carlo (SMMC) approach}

 We first discuss the SMMC method in the framework of the canonical ensemble (at fixed temperature) and then describe how to translate its results to fixed excitation energy.

\subsection{Hubbard-Stratonovich transformation for the canonical ensemble and the SMMC method}

The Gibbs operator $e^{-\beta H}$ at inverse temperature $\beta=1/T$ is also the system's propagator in imaginary time $\beta$.  The Trotter product of the propagator is obtained by dividing the time interval $[0,\beta]$ into $N_t$ time slices  of length $\Delta \beta$ each, and rewriting $e^{-\beta H}  = \left(e^{-\Delta \beta H}\right)^{N_t}$.  Expressing the two-body interaction as a quadratic form in the one-body densities, the short-time propagator $e^{-\Delta \beta H}$ can be written as a integral over auxiliary fields $\sigma$ of  one-body propagators described by a one-body Hamiltonian $h_\sigma$ that is linear in the one-body densities.  A different set of fields $\sigma(\tau_n)$ is required for each time slice $\tau_n  = n \Delta \beta \;\; (n=1,\ldots,N_t)$. The finite-time propagator is then written as a functional integral
\be\label{HS}
e^{-\beta H} = \int D[\sigma] G_\sigma U_\sigma
\ee
over time-dependent auxiliary fields $\sigma(\tau)$ with a Gaussian weight $G_\sigma$. Here $\hat U_\sigma=e^{-{ \Delta}\beta h_\sigma(\tau_{N_t})} \ldots e^{-{ \Delta}\beta h_\sigma(\tau_1)}$ is the many-particle propagator for a time-dependent one-body Hamiltonian $h_\sigma(\tau)$ that describes non-interacting particles moving in external fields $\sigma(\tau)$. Relation (\ref{HS}) is known as the Hubbard-Stratonovich (HS) transformation~\cite{HS-trans}.

The thermal expectation value of an observable $O$ is calculated from
 \be \label{observable}
\langle O\rangle = {\Tr \,( O e^{-\beta H})\over  \Tr\, (e^{-\beta H})} = {\int D[\sigma] W_\sigma \Phi_\sigma \langle O \rangle_\sigma
\over \int D[\sigma] W_\sigma \Phi_\sigma} \;,
\ee
where $W_\sigma = G_\sigma |\Tr\, U_\sigma|$ is a positive-definite function,  $\Phi_\sigma = \Tr\, U_\sigma/|\Tr\, U_\sigma|$ is the Monte Carlo sign function, and $\langle O \rangle_\sigma\equiv
 {\rm Tr} \,( O U_\sigma)/ {\rm Tr}\,U_\sigma$ is the thermal expectation value of the observable at a given configuration of the auxiliary fields $\sigma$.  In SMMC, we use a Metropolis algorithm to choose configurations $\sigma_k$ that are distributed according to the weight function $W_\sigma$. The thermal expectation value in (\ref{observable}) is then estimated from
\be
\langle O\rangle \approx { \sum_k
  \langle  O \rangle_{\sigma_k} \Phi_{\sigma_k} \over \sum_k \Phi_{\sigma_k}} \;.
\ee

In the finite nucleus, it is necessary to calculate observables at fixed number of protons $Z$ and neutrons $N$. To that end, we use the canonical ensemble in which the traces in (\ref{observable}) and in the weight function $W_\sigma$ are calculated at fixed $Z$ and $N$.  The projection on a fixed number of particles can be expressed as a Fourier sum. For example, for $N_s$ single-particle orbitals, the canonical partition function $\Tr_A\, U_\sigma$ for $A$ particles is given by
\begin{eqnarray}\label{canonical}
\Tr_A U_\sigma =\frac{e^{-\beta\mu  A}}{N_s}
\sum_{m=1}^{N_s} e^{-i\phi_m A}\,\Tr\, \left( e^{\beta\mu \hat A} e^{i\phi_m\hat A} U_\sigma \right)
\;,
\end{eqnarray}
where $\hat A$ is the particle-number operator, $\phi_m=2\pi m/N_s \;\; (m=0,\ldots,N_s)$ are quadrature points and $\mu$ is a real chemical potential [used to stabilize numerically the sum in (\ref{canonical})]. The trace on the right-hand side of (\ref{canonical}) is a grand-canonical trace evaluated over the Fock space with all values of particle number and is given by $\det \left({\bf 1}+e^{i\phi_m}e^{\beta\mu}{\bf U}_\sigma\right)$, where ${\bf U}_\sigma$ is the $N_s\times N_s$  matrix representing $U_\sigma$ in the single-particle space.

The SMMC method is particularly useful in the microscopic calculations of statistical properties of nuclei~\cite{SMMC-statistical}.  Many of the early applications were to medium-mass nuclei, and more recently the method was extended to heavy nuclei, overcoming a number of technical challenges~\cite{Alhassid2008}.

\subsection{Microcanonical ensemble}

In the SMMC method we calculate observables at fixed temperature, i.e., in the canonical ensemble. However, nuclear properties are usually measured at a fixed energy $E$, necessitating the use of the microcanonical ensemble $\delta(E -H)$.  The microcanonical ensemble is related to the canonical ensemble by an inverse Laplace transform
\be \label{micro}
\delta(E -H) = \int_{-i\infty}^{i\infty} d\beta e^{\beta E} e^{-\beta H} \;.
\ee
The inverse Laplace transform is numerically unstable but can be evaluated in the saddle-point approximation.

An example is the level density $\rho(E)$ at energy $E$ whose relation to the canonical partition function is obtained by taking the trace of Eq.~(\ref{micro})
\be\label{level-density}
\rho(E) = \int_{-\infty}^{\infty} d\beta e^{\beta E} Z(\beta) \;.
\ee
The average level density is obtained by evaluating (\ref{level-density}) in the saddle-point approximation. The saddle-point condition $-\partial \ln Z/\partial \beta=E$ determines $\beta$ as a function of $E$ and the level density is approximated by
\be
\rho(E) \approx {1 \over \sqrt{2\pi T^2 C}} e^{S(E)} \;,
\ee
where  $S(E)$ and $C$  are the canonical entropy and heat capacity, respectively.
In SMMC, we calculate the energy $E=E(\beta)$ as a function of $\beta$ and  integrate the find the partition function, i.e., $\ln Z(\beta) = \ln Z(0) -\int_0^\beta d\beta' E(\beta')$. The entropy is then calculated from $S(E) = \ln Z(\beta) + \beta E$, while the heat capacity is found by taking a derivative $C=dE/dT$. The latter derivative has large statistical errors at low $T$. These errors can be reduced significantly by calculating the derivative inside the HS integral and taking into account correlated errors~\cite{Liu2001}.

\section{Emergence of collectivity in the shell model approach}

Heavy nuclei are known to exhibit various types of collectivity, e.g., vibrational collectivity in spherical nuclei and rotational collectivity in well-deformed nuclei. Models such as the geometric Bohr Hamiltonian~\cite{Bohr} and the interacting boson model~\cite{IBM} have been successful in describing these types of collectivity. However, a microscopic description in the framework of a truncated spherical shell model has been a major challenge. In particular, heavy nuclei can have large deformation that are more difficult to reproduce in a truncated spherical model space.

The dimensionality of the many-particle model space that is required to describe such heavy deformed nuclei is many orders of magnitude beyond the capability of conventional diagonalization methods~\cite{Alhassid2008}. The SMMC method enables calculations within such large model spaces. However, SMMC is a suitable method for calculating thermal and ground-state observables, but not for calculating detailed level schemes. This poses a new challenge in the framework of SMMC since the specific type of collectivity is usually identified through its signatures in spectroscopy.

This difficulty can be overcome by calculating thermal observables that are sensitive to the specific type of collectivity.  We have identified such a thermal observable in $\langle {\bf J}^2\rangle_T$, where ${\bf J}$ is the total angular momentum of the nucleus~\cite{Alhassid2008,Ozen2012}.  At low temperatures $T$ this observable is dominated by the ground-state band and for an even-even nucleus~\cite{Fang2005}
\begin{eqnarray}\label{J2-theory}
\langle \mathbf{J}^2 \rangle_T \approx
 \left\{ \begin{array}{cc}
 30 { e^{-E_{2^+}/T} \over \left(1-e^{- E_{2^+}/T}\right)^2} &{\rm vibrational\; band}  \\
 \frac{6}{E_{2^+}} T & {\rm rotational \;band}
 \end{array} \right.
\end{eqnarray}
where $E_{2^+}$ is the excitation energy of the first $2^+$ level. Thus in the limit of sufficiently low temperatures a rotational nucleus is characterized by the linear temperature dependence of $\langle {\bf J}^2\rangle_T$, in contrast with a vibrational nucleus for which $\langle {\bf J}^2\rangle_T$  exhibits a non-linear dependence on temperature.

\subsection{Rotational collectivity in heavy deformed nuclei}

\begin{figure}[]
  \includegraphics[width=0.9\textwidth]{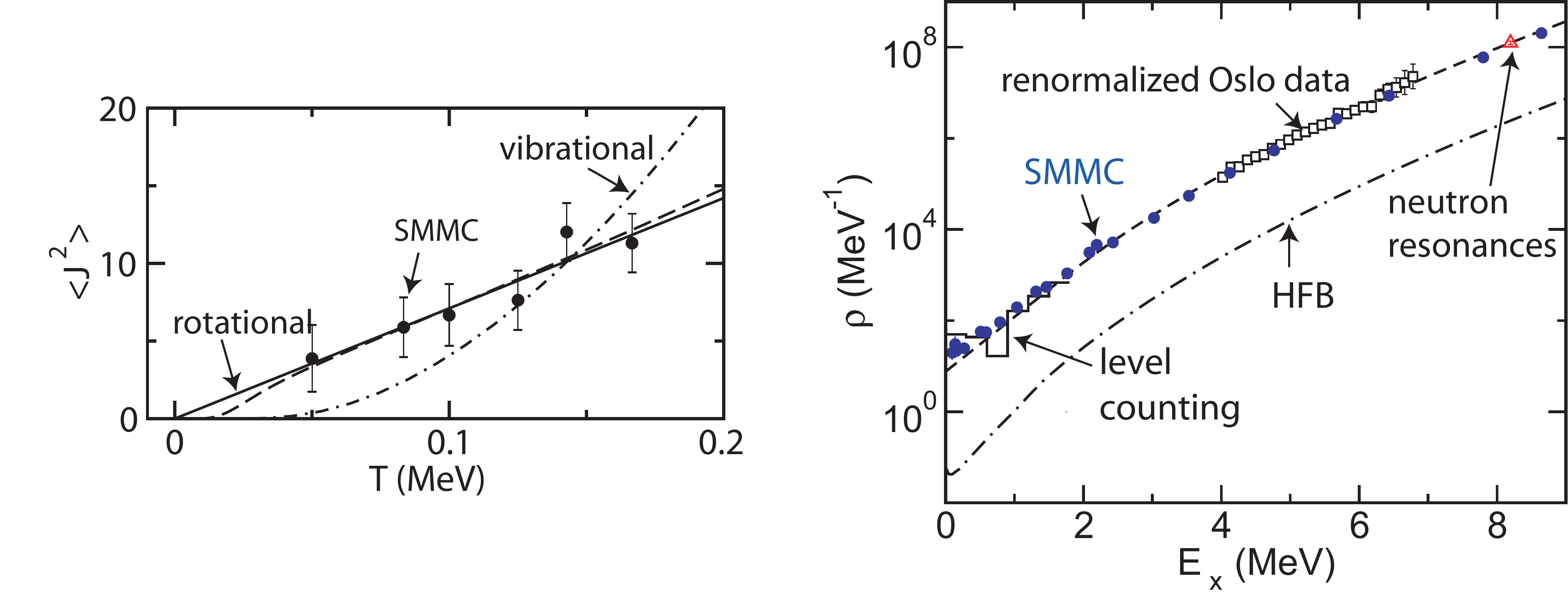}
  \caption{Left: $\langle {\bf J}^2\rangle_T$ as a function of temperature $T$ for $^{162}$Dy. The solid circles with statistical errors are the SMMC results. The solid line is a fit to the rotational band model and the dash-dotted line is a fit to the vibrational band model [see Eq.~(\ref{J2-theory})].  Right: total state density $\rho$ of $^{162}$Dy versus excitation energy $E_x$. The SMMC state density (solid circles) is compared with experimental results: level counting at low energies (histograms), neutron resonance data (triangle) and data measured by the Oslo method~\cite{Oslo} (open squares).  Adapted from Ref.~\cite{Alhassid2008}.}
\label{J2-level-dy162}
\end{figure}

We first discuss the example of a strongly deformed rare-earth nucleus, $^{162}$Dy~\cite{Alhassid2008}. The left panel of Fig.~\ref{J2-level-dy162} shows $\langle {\bf J}^2\rangle_T$ as a function of temperature $T$. The solid circles are the SMMC results and the vertical bars are the statistical errors. The solid and dash-dotted lines are, respectively, fits to the rotational- and vibrational-band formulas in Eq.~(\ref{J2-theory}). The agreement of the straight line fit with the SMMC results confirms the rotational character of $^{162}$Dy. Furthermore, from the slope of the straight line we can determine the ground-state moment of inertia $I_{\rm g.s.} =35.5 \pm 3.3$ MeV$^{-1}$, in agreement with the experimental value of $I_{g.s.} = 3/E_{2^+}= 37.2$ MeV$^{-1}$.

The total state density of $^{162}$Dy is shown in the right panel Fig.~\ref{J2-level-dy162} as a function of excitation energy $E_x$. The SMMC state density (solid circles) compares well with various experimental results. The dash-dotted line is the level density obtained in the Hartree-Fock-Bogoliubov (HFB) approximation. The SMMC density is enhanced when compared with the HFB density since it includes rotational bands that are built on top of intrinsic band heads.

\subsection{Transition from vibrational to rotational collectivity in heavy nuclei}

We have studied families of even-even samarium  ($^{148-154}$Sm) and neodymium ($^{144-152}$Nd) isotopes~\cite{Ozen2012}. These families are known to exhibit shape transitions from spherical to deformed nuclei as the number of neutrons increases from shell closure towards the mid-shell region. In the HFB approximation we find that $^{148}$Sm is spherical in its ground state (i.e., at $T=0$). $^{150}$Sm acquires a small deformation in its ground state while $^{154}$Sm is a well-deformed nucleus. $^{152}$Sm is known as an $X(5)$ nucleus defining the phase transition between spherical and axially-deformed nuclei~\cite{Iachello2001,Casten2001}. Similarly $^{144}$Nd and $^{146}$Nd are spherical while $^{148}$Nd acquires a small deformation that increases with the number of neutrons.

\begin{figure}[]
  \includegraphics[height=0.21\textheight]{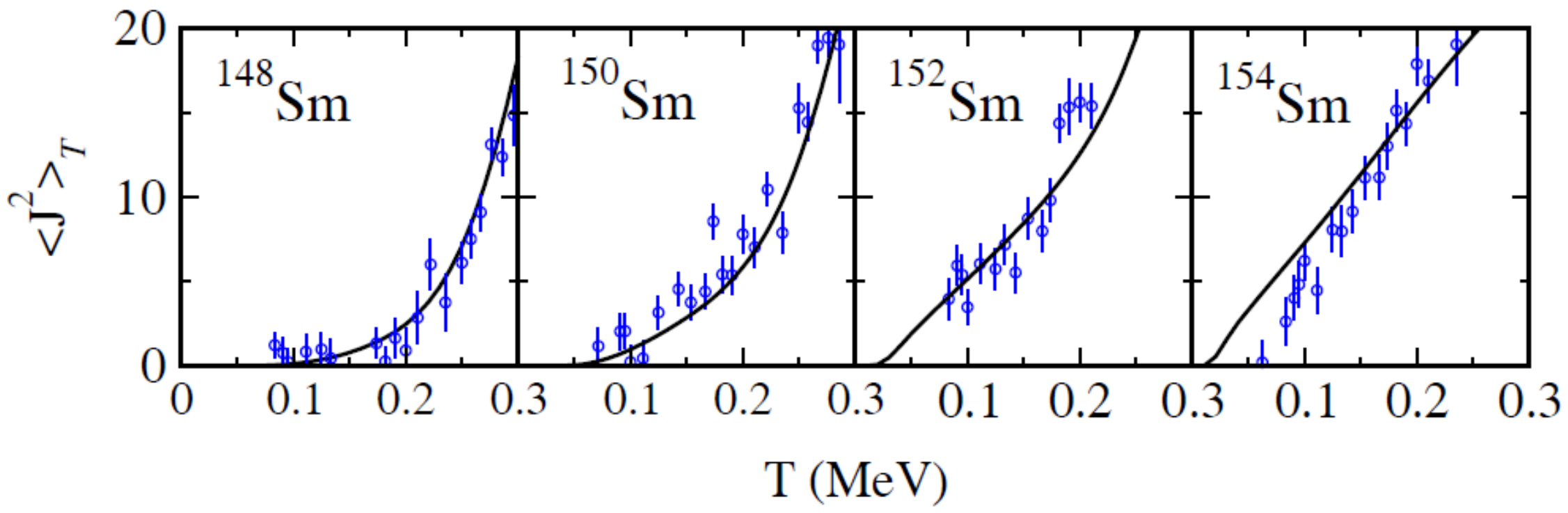}
  \caption{$\langle {\bf J}^2\rangle_T$ as a function of temperature $T$ for a family of even samarium isotopes $^{148-154}$Sm. The circles with error bars are the SMMC results and the solid lines are obtained from Eq.~(\ref{J2-exp}) using a complete set of low-lying experimental energy levels and an experimentally determined BBF level density (except for $^{154}$Sm for which only discrete levels are used). Adapted from Ref.~\cite{Ozen2012}.}
\label{Sm-J2}
\end{figure}

Fig.~\ref{Sm-J2} shows $\langle {\bf J}^2\rangle_T$ versus temperature $T$ for the samarium isotopes.  We can clearly identify in the behavior of $\langle {\bf J}^2\rangle_T$ a crossover from the vibrational nucleus $^{148}$Sm to the rotational nucleus $^{154}$Sm. The solid lines are extracted from experimental data using
\be\label{J2-exp}
 \langle \mathbf{J}^2 \rangle_T  =  \frac{1}{Z(T)} \left(\sum_i^N J_i(J_i+1)(2J_i+1)e^{-E_{i}/T} +   \int_{E_{N}}^\infty d E_x \: \rho(E_x) \: \langle \mathbf{J}^2 \rangle_{E_x} \; e^{-E_x/T} \right),
\ee
where $Z(T)=\sum_{i}^{N} (2J_i+1) e^{-E_i/T} + \int_{E_{N}}^\infty d E_x \rho(E_x) e^{-E_x/T}$ is the corresponding  partition function. The energies $E_i$ define a complete set of low-lying experimental energy levels with spin $J_i$, while $\rho(E_x)$ is a backshifted Bethe formula (BBF) level density extracted from experimental data (i.e, level counting at low energies and neutron resonance data at the neutron threshold energy).  The BBF level density is used above a certain excitation energy $E_N$ below which a complete set of experimental discrete levels is known. We observe an overall good agreement between the SMMC and experimental results.

\subsection{Phase transitions and the collective enhancement factors}

\begin{figure}[]
  \includegraphics[width=0.95\textwidth]{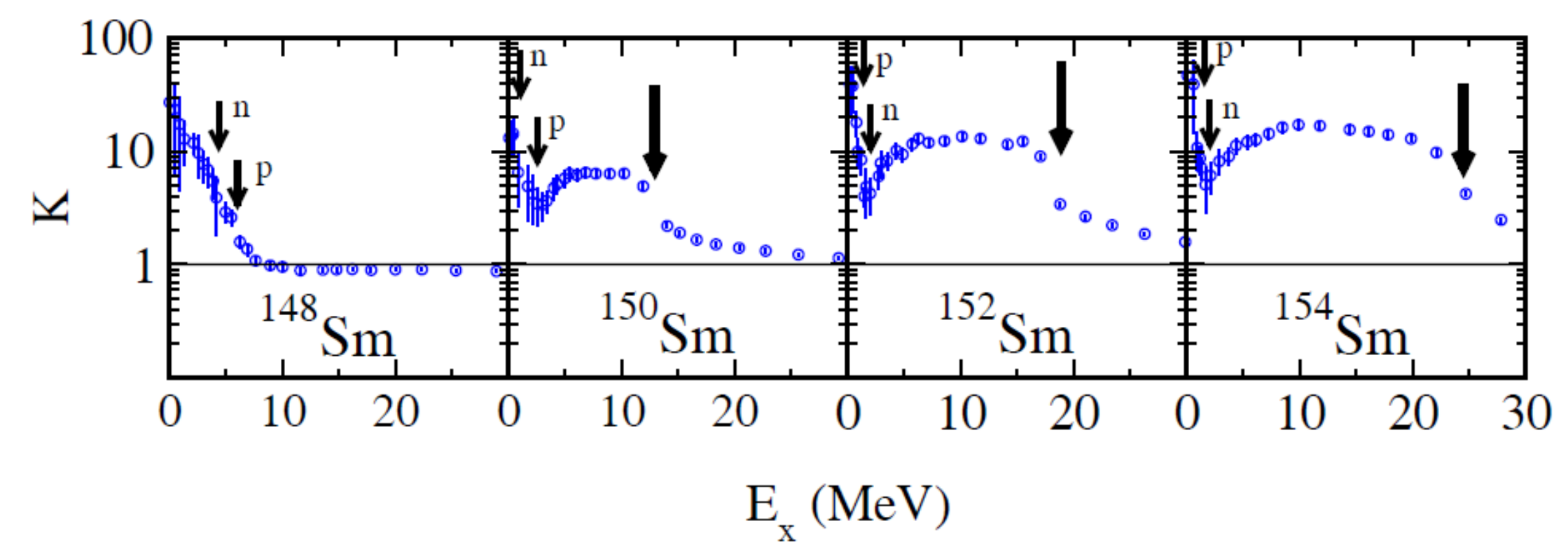}
  \caption{The enhancement factor $K$ as a function of excitation energy $E_x$ in even samarium isotopes $^{148-154}$Sm. The thin arrows indicate the excitation energies of the proton and neutron pairing transitions and the thick arrows correspond to the shape transitions. Adapted from Ref.~\cite{Ozen2012}.}
\label{Sm-enhancement}
\end{figure}

Collective states lead to enhancement of level densities. Most available expressions for the collective enhancement factors are phenomenological~\cite{RIPL}. The level density calculated in the finite-temperature HFB approximation only counts the intrinsic states. Therefore, we can define collective enhancement factor by $K = \rho_{\rm SMMC}/\rho_{\rm HFB}$, measuring the ratio between the SMMC total state density and the intrinsic HFB density.  In Fig.~\ref{Sm-enhancement} we show the calculated $K$ (on a logarithmic scale) as a function of excitation energy $E_x$ for the family of samarium isotopes $^{148-154}$Sm.

In the HFB approximation we can identify pairing and shape phase transitions which are indicated by arrows in Fig.~\ref{Sm-enhancement}. There are two pairing transitions, one for protons and a second for neutrons. $^{148}$Sm is spherical in its ground state so that only the pairing transitions occur. In such a spherical nucleus the collective enhancement originates in vibrational states, and we observe that $K$ decays to $\sim 1$ above the pairing transitions. The heavier samarium isotopes shown in Fig.~\ref{Sm-enhancement} are all deformed in their ground state and $K$ has a minimum above the pairing transitions. These observations suggest that the decay of vibrational collectivity is correlated with the disappearance of pairing correlations.  In the deformed samarium isotopes, we observe that as the excitation energy continues to increase $K$ increases and reaches a plateau because of the contribution from rotational collective states before it decays in the vicinity of the shape phase transition energy. This behavior indicates that the rotational enhancement factor is correlated with the shape transition. This is reasonable since a spherical nucleus can no longer support rotational bands.

\section{Conclusion}

The microscopic calculations of statistical properties of nuclei at finite excitation energies in the framework of the configuration-interaction shell model approach require very large model spaces. Such calculations have become possible using the SMMC method.  We have extended SMMC to heavy nuclei and identified thermal signatures of the transition from vibrational to rotational collectivity in families of rare-earth isotopes.

We have found that the damping of vibrational and rotational collectivity as a function of excitation energy correlates with the pairing and shape phase transitions, respectively. Thus, the collective enhancement factor of the level density exhibits signatures of these phase transitions as a function of excitation energy.

\begin{theacknowledgments}
It is a great pleasure for us to dedicate this article to Franco Iachello on the occasion of his 70th birthday and in appreciation of his inspiring and fruitful collaborations with Y.A.
This work was supported in part by the DOE grant DE-FG-0291-ER-40608, and by the JSPS Grant-in-Aid for Scientific Research (C) No.~22540266. Computational cycles were provided by the NERSC and  Yale University High Performance Computing Center.

\end{theacknowledgments}

\end{document}